\documentclass[showpacs,amsmath,reprint,aps]{revtex4-1}

\usepackage{graphicx}
\usepackage{hyperref}
\begin{document}

\title{Gate-voltage induced trions in suspended carbon nanotubes}
\author{M.~Yoshida}
\author{A.~Popert}
\author{Y.~K.~Kato}
\email[Corresponding author: ]{ykato@sogo.t.u-tokyo.ac.jp}
\affiliation{Institute of Engineering Innovation, The University of Tokyo, Tokyo 113-8656, Japan}

\begin{abstract}
We observe trion emission from suspended carbon nanotubes where carriers are introduced electrostatically using field-effect transistor structures. The trion peak emerges below the $E_{11}$ emission energy at gate voltages that coincide with the onset of bright exciton quenching. By investigating nanotubes with various chiralities, we verify that the energy separation between the bright exciton peak and the trion peak becomes smaller for larger diameter tubes. Trion binding energies that are significantly larger compared to surfactant-wrapped carbon nanotubes are obtained, and the difference is attributed to the reduced dielectric screening in suspended tubes.
\end{abstract}
\pacs{78.67.Ch, 71.35.Pq, 78.55.-m, 85.35.Kt}

\maketitle

The quasi-one-dimensional geometry of single-walled carbon nanotubes (SWCNTs) results in enhanced Coulombic effects that are sensitive to environmental dielectric screening.  The strong electron-hole attractive force generates tightly bound excitons, but their binding energies can differ by a few hundred meV for surfactant-wrapped nanotubes compared to suspended nanotubes \cite{Dukovic:2005,Lefebvre:2008}. Also because of the limited screening, a charged carrier can be bound to an exciton to form a trion that is stable even at room temperature \cite{Matsunaga:2011, Santos:2011}. Trions provide additional degrees of freedom for manipulating the optical properties of SWCNTs as they possess both charge and spin, and consequently they have received considerable attention. Various techniques including photoluminescence (PL) \cite{Park:2012,Mouri:2013,Akizuki:2014}, electroluminescence \cite{Jakubka:2014}, absorption \cite{Hartleb:2015}, and time-resolved \cite{Koyama:2013, Nishihara:2013, Yuma:2013} spectroscopy have been used to investigate trions, and it has been shown that trion generation in micelle-wrapped SWCNTs involves trapped charges \cite{Santos:2011,Mouri:2013,Yuma:2013}. Similar to the case of excitons, significantly enhanced trion binding energies are expected in suspended SWCNTs \cite{Ronnow:2010, Watanabe:2012, Bondarev:2014}, but it has been pointed out that trion formation is difficult due to low exciton-carrier scattering rates \cite{Konabe:2012}.

Here we report on gate-voltage induced trions in suspended carbon nanotubes within field-effect transistor structures. When a gate voltage is applied, a peak emerges at an energy below the $E_{11}$ bright exciton emission. Excitation spectroscopy under the application of the gate voltage shows that the absorption resonances for both emission peaks are the same, indicating that they arise from the same nanotube. Gate-voltage dependence measurements reveal that bright exciton quenching and the appearance of the lower energy peak occur at the same voltage, consistent with the picture that electrostatic doping leads to trion formation. This interpretation is further confirmed by the observation that the energy separation between the bright exciton peak and the gate-induced peak becomes smaller for larger diameter tubes. We obtain trion binding energies that are considerably larger compared to surfactant-wrapped tubes, as expected from the dielectric screening effects.

\begin{figure}
\includegraphics{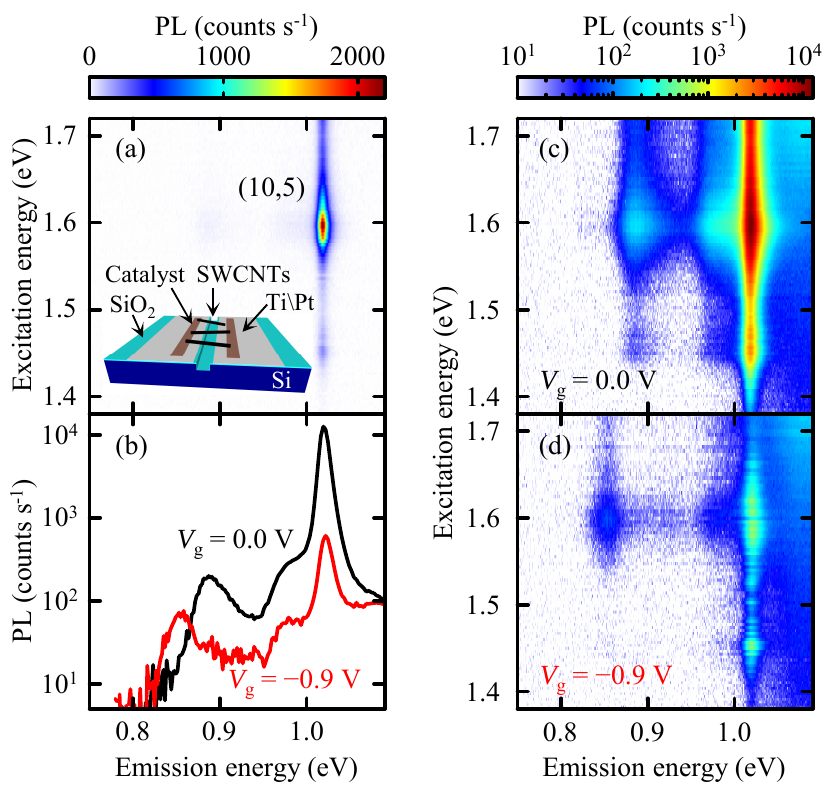}
\caption{\label{fig1}
(Color)
(a)~A PLE map of an individual nanotube in a device taken with $P=5$~$\mu$W. The absorption peaks at 1.60 and 1.45~eV correspond to the $E_{22}$ and $E_{11}^1$ resonances, respectively \cite{Lefebvre:2008}. The inset is a device schematic.
(b)~PL spectra taken at $V_\text{g}=0.0$~V (black) and $V_\text{g}=-0.9$~V (red). An excitation energy of 1.60~eV is used.
(c) and (d)~PLE maps at $V_\text{g}=0.0$~V and $V_\text{g}=-0.9$~V, respectively. For (b-d), the excitation power is fixed at $P=300$~$\mu$W, and the laser polarization is parallel to the nanotube axis. Weak emission above 1.0~eV comes from Si.
}\end{figure}

Our suspended SWCNT field-effect transistors \cite{Yasukochi:2011, Kumamoto:2014, Yoshida:2014} are fabricated from $p$-Si substrates with a resistivity of $15\pm5$ m$\Omega\cdot$cm and 100-nm-thick oxide [Fig.~\ref{fig1}(a) inset]. We form 500-nm-deep trenches with a width of 1~$\mu$m by electron beam lithography and dry etching. The samples are then oxidized in an annealing furnace at $900^{\circ}$C for an hour in order to form a 20-nm-thick SiO$_2$ layer in the trenches. Another electron beam lithography step defines the electrode patterns, and we deposit Ti (2~nm)$\backslash$Pt (20~nm) using electron beam evaporation. Following a lift-off process, catalyst regions near the trenches are patterned by a third electron beam lithography step. In order to grow SWCNTs with a wide diameter range, we use either Co/silica dissolved in ethanol \cite{Yoshida:2014} or a 0.2-nm thermally evaporated Fe film \cite{Inoue:2013}. After depositing and lifting off the catalyst, the samples are heated in air for 5~min at $400^{\circ}$C. We note that the Co/silica catalyst is placed on top of the Pt electrodes, while the Fe catalyst covers the SiO$_2$ area in small windows opened up within the electrodes. Finally, SWCNTs are synthesized by chemical vapor deposition \cite{Imamura:2013} at $800^{\circ}$C with a growth time of 1~min.

PL measurements are performed with a home-built laser-scanning confocal microscope system \cite{Moritsubo:2010, Watahiki:2012, Yokoyama:2014}.  In order to reduce gate hysteresis \cite{Kim:2003}, samples are mounted in a vacuum chamber with optical and electrical access. For excitation, a continuous-wave Ti:sapphire laser with a wavelength range of 700 to 950~nm is focused onto the sample by an objective lens with a correction collar, a numerical aperture of 0.65, and a focal length of 3.6~mm. Emission from the samples is focused into a 300-mm spectrometer by a 50-mm focal-length lens, and the PL is dispersed using a 150~lines/mm grating blazed at 1.25~$\mu$m. PL spectra are detected by an InGaAs photodiode array with a detection wavelength from 800 to 1700~nm, or an extended range InGaAs photodiode array which can detect emission between 1000 and 2200~nm at the cost of a reduced signal-to-noise ratio. All measurements are done at room temperature.

We first perform PL excitation (PLE) spectroscopy on the suspended nanotubes to determine their chirality. Typical data taken with an excitation power $P=5$~$\mu$W is shown in Fig.~\ref{fig1}(a), where the tube chirality $(n,m)$ is identified to be (10,5) by utilizing an empirical table \cite{Ishii:2015}. We verify that the nanotube is fully suspended over the trench by taking PL and reflectivity images, and the laser polarization dependence of PL is used to check that the nanotube is relatively straight \cite{Moritsubo:2010}. After such careful characterization, we have investigated the effects of the voltage $V_\text{g}$ applied to the back gate while grounding the nanotube contacts. At $V_\text{g}=0.0$~V [Fig.~\ref{fig1}(b), black curve], the $E_{11}$ bright exciton peak is visible as well as a side peak at a lower energy, which is the $K$-momentum dark exciton state \cite{Matsunaga:2010}. For the case of $V_\text{g}=-0.9$~V [Fig.~\ref{fig1}(b), red curve], $E_{11}$ and $K$-momentum exciton peaks show quenching as a result of electrostatic doping \cite{Yasukochi:2011}, and a new redshifted peak with 168-meV energy separation from $E_{11}$ is observed. To rule out the presence of another tube in the vicinity, we have performed PLE spectroscopy at $V_\text{g}=0.0$ [Fig.~\ref{fig1}(c)] and $-0.9$~V [Fig.~\ref{fig1}(d)]. We find that the $E_{11}$ peak and the new peak exhibit similar absorption behaviors, confirming that the gate-induced peak originates from the same nanotube.

\begin{figure}
\includegraphics{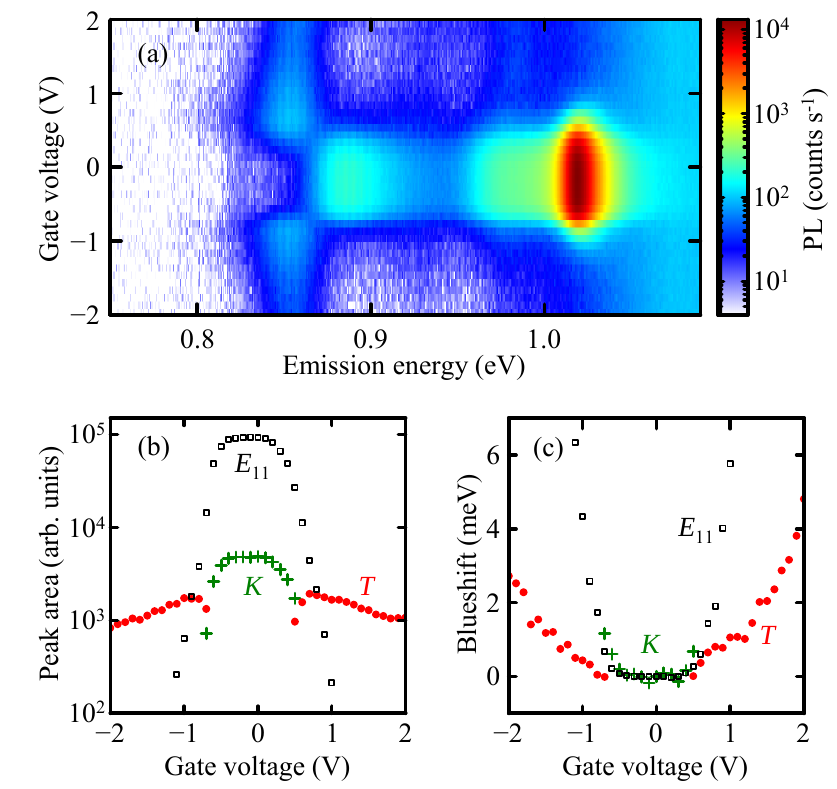}
\caption{\label{fig2}
(Color)
(a)~PL spectra as a function of gate voltage taken at a sweep rate of 40~mV s$^{-1}$. The excitation energy is 1.60~eV and $P=300$~$\mu$W is used. The laser polarization is parallel to the nanotube axis.
(b) and (c)~Gate voltage dependence of peak area and blueshift, respectively, for $E_{11}$ bright exciton (black squares, $E_{11}$), $K$-momentum dark exciton (green crosses, $K$), and gate-induced (red dots, $T$). Fitting is done by a Lorentzian plus a linear function. For~(c), the blueshifts are measured from 1020.4, 888.8, and 853.4~meV for the $E_{11}$ exciton, $K$-momentum exciton, and gate-induced peaks, respectively.
}\end{figure}

In order to characterize the gate-induced peak in detail, gate voltage dependence of PL on this nanotube has been investigated [Fig.~\ref{fig2}(a)]. By performing fits to these spectra by a Lorentzian function with a linear background, we have extracted the peak areas and emission energies for the $E_{11}$ bright exciton, the $K$-momentum exciton, and the gate-induced peaks. In Fig.~\ref{fig2}(b), the gate-voltage dependence of the spectrally integrated PL intensities of the peaks is shown, where the $E_{11}$ emission quenches exponentially with $V_\text{g}$ as observed previously \cite{Yasukochi:2011}. In contrast, the new peak emerges with an application of gate voltage, and becomes maximized at $V_\text{g}=-0.9$ and 0.7~V. Such a behavior indicates that electrostatic carrier doping gives rise to the new peak, and we therefore assign the gate-induced peak to trion emission. 

We estimate the carrier density at maximum trion intensity to be 0.014~nm$^{-1}$ by modeling the device \cite{Yasukochi:2011}. In comparison, typical chemical doping concentrations used to observe trion emission \cite{Matsunaga:2011} correspond to about an order of magnitude higher hole densities \cite{Mouri:2012}. We also find that the trion peak behaves symmetrically for positive and negative gate voltages. This results from effective masses of electrons and holes being almost the same, and have also been observed in previous work utilizing electrochemical \cite{Park:2012, Hartleb:2015} and electrolyte gating \cite{Jakubka:2014} techniques.

\begin{figure*}
\includegraphics{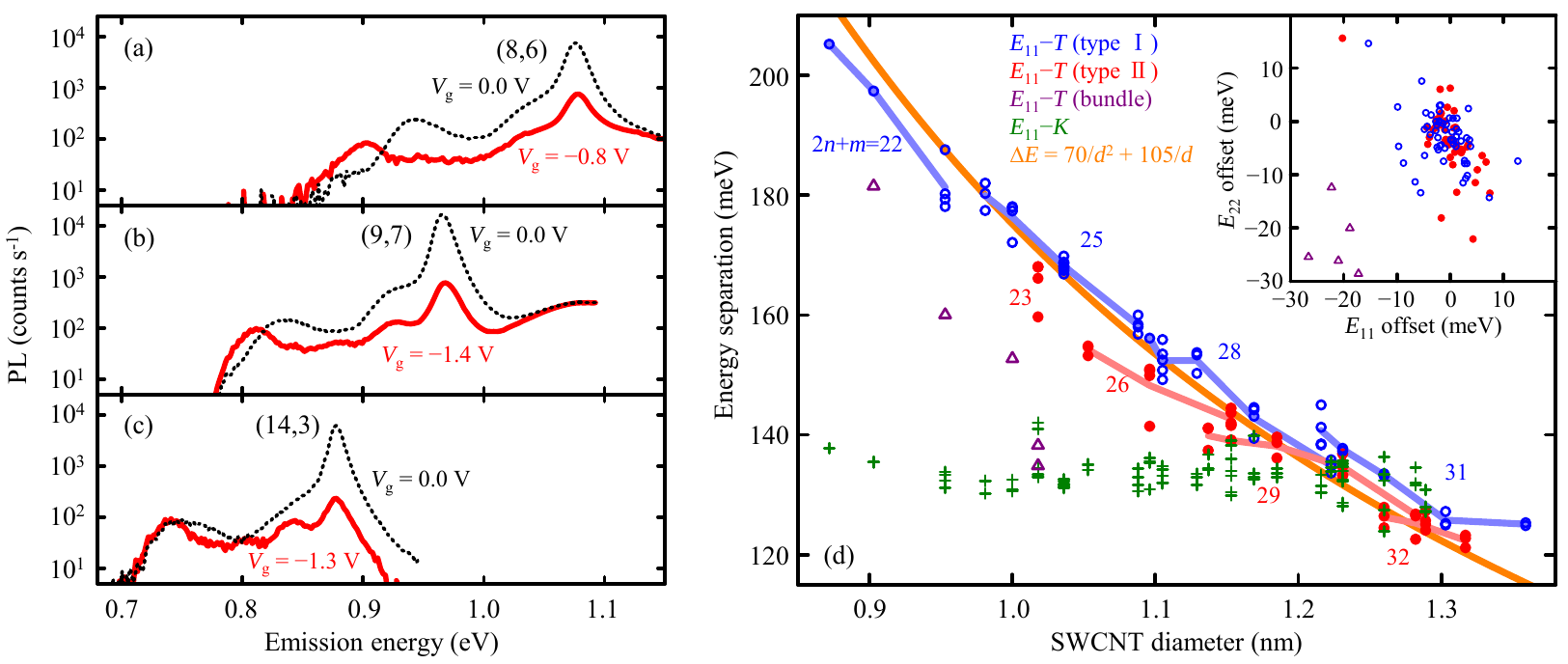}
\caption{\label{fig3}
(Color)
(a-c)~PL spectra for (8,6), (9,7), and (14,3) nanotubes, respectively. Data are taken at zero gate voltage (black broken lines) and under an application of gate voltages (red solid lines).
(d)~Nanotube diameter dependence of energy separation $\Delta E$ between $E_{11}$ and the trion peaks on 100 individual SWCNTs. Blue circles and red dots correspond to type~I and type~II nanotubes, respectively. Purple triangles show $\Delta E$ for bundled tubes. Green crosses indicate data on energy splitting between $E_{11}$ and $K$-momentum exciton peaks. An orange line is a fit, while blue and red lines show the family patterns. All of the data are taken at $E_{22}$ excitation and $P=300$~$\mu$W. Laser polarization is parallel to each nanotube axis. Inset shows $E_{11}$ and $E_{22}$ energy offsets from tabulated PLE data in Ref.~[\citenum{Ishii:2015}].
}\end{figure*}

As the gate voltage increases further, the intensity of the trion peak gradually decreases [Fig.~\ref{fig2}(b)], which may be caused by a reduction of $E_{22}$ absorption cross section or changes in Auger recombination rates. In addition, the emission energies for all of the peaks show slight blueshifts with increasing gate voltage [Fig.~\ref{fig2}(c)]. Such gate-induced blueshifts for $E_{11}$ bright exciton and trion peaks have been observed previously \cite{Yasukochi:2011, Jakubka:2014}, and may be caused by a combination of band-gap renormalization, band filling, and changes in the binding energies \cite{Spataru:2010}. 

Now we investigate the diameter dependence in order to determine the trion binding energy in suspended nanotubes. In Figs.~\ref{fig3}(a-c), we present PL spectra taken with and without gate voltages on three nanotubes of different chiralities. All of the nanotubes show the emergence of the trion peak as well as quenching of the $E_{11}$ and $K$-momentum exciton peaks upon application of gate voltages. Furthermore, the energy separation $\Delta E$ between the $E_{11}$ peaks and the trion peaks decreases as the tube diameter becomes larger.

Such measurements are systematically performed on 100 individual nanotubes to obtain the nanotube diameter dependence of $\Delta E$ [blue circles and red dots in Fig.~\ref{fig3}(d)]. We note that $\Delta E$ has been measured at a gate voltage where the $E_{11}$ intensity has quenched by a factor of 10 to 100, such that both $E_{11}$ and trion peak positions can be accurately determined. A clear diameter dependence as well as family patterns are observed, similar to the diameter dependence of trions in surfactant-wrapped SWCNTs \cite{Matsunaga:2011, Santos:2011, Park:2012}. In comparison, the  $K$-momentum exciton peaks [green crosses in Fig.~\ref{fig3}(d)] do not show such a dependence \cite{Matsunaga:2010}.

The diameter dependence of $\Delta E$ can be used to extract the trion binding energy. As the trion emission originates from a bound state of a triplet exciton and a charged carrier, the $E_{11}$-trion splitting is given by $\Delta E=A/d+B/d^2$, where the first term represents the trion binding energy and the second term is the singlet-triplet splitting \cite{Matsunaga:2011, Santos:2011, Park:2012}. Using $B=70$~meV$\cdot$nm$^2$ for suspended nanotubes \cite{Matsunaga:2010}, we obtain $A=105$~meV$\cdot$nm by fitting the data [orange line in Fig.~\ref{fig3}(d)]. The value is significantly larger than the binding energy in surfactant-wrapped SWCNTs which ranges from 49 to 85~meV$\cdot$nm \cite{Matsunaga:2011, Santos:2011, Park:2012}, and the difference is likely caused by dielectric screening from the surrounding ambient \cite{Ronnow:2010, Watanabe:2012, Bondarev:2014}. As the exciton binding energy in SWCNTs increases by a factor of 1.5 for suspended tubes compared to surfactant-wrapped tubes \cite{Lefebvre:2008}, the extracted value of $A$ is fairly reasonable since the power-law scaling of exciton and trion binding energies with respect to dielectric constant is similar \cite{Perebeinos:2004,Ronnow:2010}.

We note that 5~suspended SWCNTs [triangles in Fig.~\ref{fig3}(d)] display $\Delta E$ that is lower by about 30~meV from the fit.
For these tubes, both $E_{11}$ and $E_{22}$ energies are significantly redshifted compared to tabulated PLE data in Ref.~[\citenum{Ishii:2015}] as shown in Fig.~\ref{fig3}(d) inset. The redshifted peaks forming satellite spots in the PLE map have been interpreted as bundles of nanotubes with the same chirality \cite{Ishii:2015}, in which the increased screening reduces the optical transition energies with respect to individual SWCNTs. The smaller $E_{11}$-trion energy splitting is consistent with this interpretation.

The observation of trions in electrostatically-doped clean SWCNTs opens up new prospects for spin manipulation.
The trions in suspended nanotubes may be delocalized, in which case they would allow for spin transport along the tube. As trions are charged particles, it should be possible to drive them by applying longitudinal electric fields. There may be new opportunities for investigating spin properties in nanotubes by utilizing trion emission for detection.

In summary, we have investigated gate-induced trion emission in suspended carbon nanotubes with assigned chirality. Gate voltage dependence of PL shows coincidence of bright exciton quenching and gate-induced peak emergence, indicating that trions are generated by electrostatic doping. By performing measurements on tubes with different chiralities, we confirm that the $E_{11}$-trion energy splitting becomes smaller with increasing tube diameter. The trion binding energies are found to be significantly larger than those in surfactant-wrapped SWCNTs, which can be explained by dielectric constant scaling of the binding energies.

\begin{acknowledgments}
We thank T. Kan and I. Shimoyama for the use of the electron beam evaporator, S. Chiashi and S. Maruyama for the thermal evaporator, and S. Yamamoto for the plasma etcher. We also acknowledge Y. Ueda for technical assistance. Work supported by JSPS (KAKENHI 24340066), the Canon Foundation, the Sasakawa Scientific Research Grant, and MEXT (Photon Frontier Network Program, Nanotechnology Platform). The devices were fabricated at the Center for Nano Lithography \& Analysis at The University of Tokyo. M.Y. is supported by ALPS.
\end{acknowledgments}

\end{document}